# Calculation of Ar photoelectron satellites in hard X-ray region

V. G. Yarzhemsky[1, 2], M. Ya Amusia[3, 4]

[1] N. S. Kurnakov Institute of General and Inorganic Chemistry, 119991 Moscow, Russia
[2] Moscow Institute of Physics and Technology, Dolgoprudny, 141700, Moscow Region, Russia
[3] Racah Institute of Physics, the Hebrew University, Jerusalem 91904, Israel
[4] A F Ioffe Physical-Technical Institute, St Petersburg 194021, Russia

The intensities of photoelectron satellite lines, corresponding to double core hole (DCH) states of Ar $1s$ ionization by hard X-rays are calculated using the many-body-perturbation theory. Calculations support the interpretation of the most intense lines as the shake-up excitations $2p \rightarrow 4p$. It is demonstrated that the intensities of the spectrum lines corresponding to $4s$ (and $3d$) excited states in DCH field can be explained only taking into account the direct knock-up process $2p \rightarrow 3d$ along with shake-up process $1s \rightarrow 4s$ that accompanies $2p$ photoionization.

## I. INTRODUCTION

Recent experiments in hard X-ray photoelectron spectroscopy (HARPES) resulted in an essentially new type of photoelectron satellites, corresponding to final states with double core hole (DCH) state and one excited electron [1]. Already several years ago in [2] some of the DCH states in Ar$1s$ photoionization corresponding to excitations $2p \rightarrow 4p$ were recorded experimentally and their positions were calculated. Well resolved structure in DCH energy region [1] is the basis for theoretical calculation of electron correlation with high energy transfer, what corresponds to HARPES.

Along with strong lines, a very interesting result of [1] is the identifications of excitations $1s \rightarrow 4s$ with extremely large energy transfer to satellite and very small relative intensity, about $10^{-4}$ of the main line.

These measurements became possible due to modern HARPES technique in photoelectron spectroscopy. HARPES is versatile for the investigation of bulk electronic structure of functional materials [3, 4]. Theoretical calculations are required for better understanding of physics of this essentially new type of excitation. Due to significant energy difference from classical electron spectroscopy physical phenomena related to HARPES, e.g. non-dipolar effects and inter-channel correlations should be specially considered for this energy region [5, 6].

The satellite peaks in the photoelectron spectra appear due to many-electron correlation, and their investigation is important for understanding the many-body nature of interaction of radiation with matter.

Satellite spectra in soft X-ray region were investigated quite well [7-13]. Corresponding satellite intensities were calculated within the sudden approximation [14-17], configuration interaction method [18], and many-body perturbation theory [13, 19, 20]. The investigation of the decay of valence satellites into different continuum channels resulted in discovery of a new physical phenomena – non-uniform lineshape broadening in different decay channels [21], which was predicted theoretically [22-24].

Two satellite series $2p \rightarrow 4p, 5p$, 5p and $1s \rightarrow 4s$ identified in [1] correspond to large ~300 eV and very large ~3000 eV energy transfer are due to many electron correlation. The main aim of the present work is to calculate satellites intensities and to clarify some specific features of the satellite excitations in the case of large energy transfer. We discuss also the problem of extension of the existing many-electron photoionization theory to HARPES photon energy region and high energy transfer to satellite states. Also we calculate the energy positions of all satellite series and for strong lines generally confirm their interpretations in [1]. As to the relatively weak lines, the interpretation of at least one of them proved to be different. It is demonstrated that it is exited instead of shake-up via knock-up process.

## II. THEORY

Theoretical approaches to the double ionization are usually based on the perturbation theory, where the potential of a hole initially created by the photon is considered as a perturbation potential (see [25] and references therein). In the lowest order of many-body perturbation theory the relaxation of the hole in the core $k$ created in photoionization, is described by the excitation of particle-hole pairs, $sj^{-1}$. Photoionization with shake-up excitation of another electron is shown in Fig. 1a. Note, that for monopole shake-up process one has $i=k$.

To account for these processes the initial self-energy $\Sigma_k(E)$ of a hole is calculated according to the following formula:

$$\mathrm{Re}\,\Sigma_k(E) = \sum_{ijs} \frac{\langle kj|U|is\rangle^2}{E - \varepsilon_i - \varepsilon_j + \varepsilon_s}. \qquad (1)$$

Here $\langle kj|U|is\rangle$ are the Coulomb inter-electron interaction matrix elements with exchange, and the sum (integral) runs over all final states $i^{-1}j^{-1}s$ including the discrete excited and continuum states $s$. One-electron energies of electrons and holes are denoted as $\varepsilon_i$, and $E$ stands for the energy parameter of the initial hole $k$. Radial parts of Coulomb matrix elements and energies are calculated using the Hartree-Fock computer code [26]. Angular momentum coupling technique is applied to analytic calculations of the angular parts of Coulomb matrix elements, i.e. the so-called weight factors [25, 27-28].

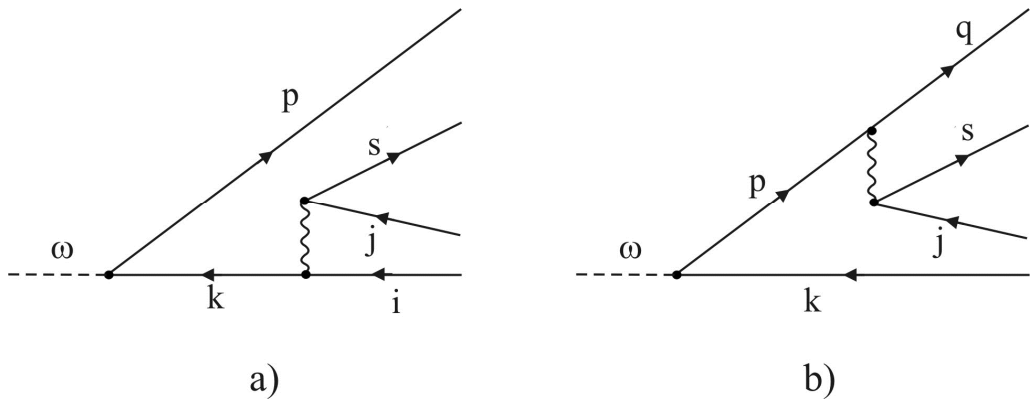

Fig. 1. Many-body theory diagrams of two channels of satellite excitation: a) excitation by core hole potential (shake-up channel) and b) excitation by outgoing electron (direct knock-up channel). A solid line with an arrow directed to the right (left) represent electrons (holes), respectively. Dashed line denotes interaction with photon and wavy line denotes for the Coulomb interaction.

The positions of lines in the spectrum are given by the solutions of the so-called Dyson equation [25]:

$$E = \varepsilon_k + \Sigma_k(E) \qquad (2)$$

The lines' intensities are proportional to the spectroscopic factors that is defined by the expression [25]:

$$f = \frac{1}{1 - \partial \Sigma_k / \partial E} . \qquad (3)$$

Here the derivatives are taken at the solutions of the equation (2). In the vicinity of the satellite under consideration, formulas (1) and (3) may be approximated by the relation, corresponding to the second order of the perturbation theory [25]:

$$f_s = \frac{\langle kj|U|ks \rangle^2}{(\varepsilon_s - \varepsilon_j)^2} \qquad (4)$$

For the excitations from valence shells the satellite intensities are about $10^{-2}$ of the main line [7-13], and the energy difference in (4) is about one a.u. If the satellite is excited from Ar 2$p$ shell, the energy denominator is about 10 a.u. and satellite intensities should be about $10^{-4}$ of the main line. This crude estimation agrees with the experimental results [1]. For the satellite excitation from Ar 1$s$ shell, whose binding energy is about 100 a.u. one can expect the values of spectroscopic factor to be about $10^{-6}$.

The same satellite states can be exited also by the outgoing photoelectron (see Fig. 1 b). The amplitude of this process is given by the following expression [13, 25]:

$$A_{kqjs}(\omega) = \int_{\varepsilon_p} \frac{D_{kp}(\omega) U_{ps,qj} d\varepsilon_p}{\omega - \varepsilon_p + \varepsilon_k + i\delta} \delta(\omega - \varepsilon_q + \varepsilon_k - \varepsilon_s + \varepsilon_j). \qquad (5)$$

In the formula (5) the energy conservation is fulfilled for the final state $q$ while the photoionization amplitude $D_{kp}(\omega)$ is calculated beyond the so-called energy conservation frame, i.e. for $\omega$ not necessary equal to $\varepsilon_p - \varepsilon_k$. Photoionization cross-sections with corresponding energy shifts were investigated in RPAE approximation [25]. It was obtained that photoionization of any level by photon energy $\omega$ with energy transfer $\Delta E_s$ to another channel may be roughly approximated by the photoionization cross-section of the same level by photons of the energy $\omega - \Delta E_s$.

This outcome of [25] can be envisaged as follows. In the shake-up channel the kinetic energy of photoelectron is $\Delta E_s$ less than in the main line channel, and the wave function with actual photoelectron energy must be used for the calculation of the dipole matrix element. This corresponds to the calculation of photoionization cross-section at photon energy $\omega - \Delta E_s$. Thus, the formula for the satellite intensity relative to any reference photoelectron line $r$ can be presented as:

$$\tilde{f}_s = \frac{f_s \sigma_k(\omega - \Delta E_s)}{\sigma_r(\omega)}. \qquad (6)$$

Here spectroscopic factors are defined by the formula (3). Note, that in [1] and in the present work the reference line is Ar 1s line, and the satellite excitations accompanying 1s- and 2p-photoionization are considered.

Photoionization cross-sections of Ar 1s and Ar 2p shells for different photon energies, calculated using the program from [29] are presented in the Table 1. Photon energy 3900 eV corresponds to the single photoionization. Photoionization cross-section of Ar 1s line at 3650 eV corresponds to the satellite excitation $2p \rightarrow 4p$. It is seen that this approach enhances the cross-section by 18%. In the case of Ar 2p line a transition from photon energy 3900eV to 700eV can be considered as a very rough approximation to the satellite excitation $1s \rightarrow 4s$ effect. Nevertheless, it is possible to conclude that in the case of satellite excitation from core shells the photoionization cross-section, which has to be used for the calculation of the satellite intensity is significantly larger then that for the main line

Table 1. Theoretical photoionization cross-sections of Ar 1s and Ar 2p shells at different photon energies (in kilobarns)

| | Cross-section | | |
|---|---|---|---|
| Shell \ photon energy | 3900 eV [a] | 3650 eV [b] | 700 eV [c] |
| Ar 1s | 48.0 | 56.8 | |
| Ar 2p | 1.84 | | 366 |

a) Actual photon energy in [1],
b) Formal photon energy corresponding to energy transfer 250 eV to the $2p \rightarrow 4p$ excitation,
c) Formal photon energy, corresponding to energy transfer 3200 eV to the $1s \rightarrow 4s$ excitation.

For the satellite excitation from valence shells one can use the photoionization cross-sections beyond the energy conservation shell [25] or neglect the difference between $\sigma_k(\omega - \Delta E_s)$ and $\sigma_k(\omega)$. On the other hand, in the case of DCH states the energy difference $\Delta E_s$ is quite significant and cannot be neglected. It is clear from the formula (5) that when a satellite is excited by the outgoing electron the main contribution to the amplitude (5) comes from the vicinity of the singularity of the integrand that corresponds to the energy conservation. In this case one can use as photoionization cross-section the corresponding value for the main line.

### III. RESULTS AND DISCUSSION

The calculations of satellites excitation energies were performed within the *LSJ* coupling scheme. The theoretical value of the relativistic 1s ionization energy 3208.9eV calculated as the difference of Dirac-Fock [30] total energies of Ar and Ar $1s^{-1}$ states is a little bit larger than the experimental ionization energy 3206.3 eV. To take into account this small systematic difference all theoretical binding energies of DCH states were shifted by the value 2.6 eV. The calculations in valence shells were performed using a non-relativistic program [26] with the *L-S* coupling for the terms. To obtain ionization energy, the non-relativistic value of the average of configuration state Ar $1s^{-1}2p^{-1}$ energy was replaced by the relativistic [30] energy Ar $1s^{-1}2p^{-1}$ averaged over $2p_{1/2}$ and $2p_{3/2}$ holes. The value 11.78 eV of the multiplet splitting between $1s^{-1}2p^{-1}(^1P)$ and $1s^{-1}2p^{-1}(^3P)$ DCH states is significantly larger than the theoretical spin-orbit splitting of 2p- shell 2.59 eV. Spin-orbit splitting for the final states was calculated using graphical representation of the spin-orbit coupling [25, 28] weight factors. Theoretical energies for the DCH states are compared with experiment in Table 2.

Table 2. Theoretical binding energies (in eV) of Ar DCH states

| State | $nl=4s^{1)}$ | $nl=4s^{3)}$ | $nl=4p^{1)}$ | $nl=4p^{2)}$ | $nl=4p^{3)}$ | $nl=5p^{1)}$ | $nl=5p^{2)}$ |
|---|---|---|---|---|---|---|---|
| $(1s^{-1}2p^{-1})(^1P_1)nl$ | 3511.9 | 3511.5 | 3514.5 | 3514.3 | 3512.8 | 3518.6 | 3519.0 |
| $(1s^{-1}2p^{-1})(^3P_0)nl$ | 3504.6 | ---- | 3504.6 | 3505.2 | 3503.3 | 3508.6 | 3509.8 |
| $(1s^{-1}2p^{-1})(^3P_1)nl$ | 3503.8 | 3501.4 | 3501.2 | 3504.0 | 3502.6 | 3507.6 | 3508.2 |
| $(1s^{-1}2p^{-1})(^3P_2)nl$ | 3499.6 | 3500.0 | 3502.0 | 3502.5 | 3500,9 | 3506.0 | 3506.6 |

1) Theoretical values, corrected by 2.6 eV (see text)
2) Experimental values (estimated from figure of [1])
3) Theoretical [2]

The agreement of theoretical binding energies with the experimental ones [1] confirms the validity of *LSJ* coupling scheme for the case under consideration.

Theoretical intensities of satellites, accompanying Ar 1*s* photoionization are compared with experimental results [1, 13] in Table 3. There is no spin-orbit splitting in the case of a singlet channel $^1P_1$, and standard formulas for satellite intensities in *L-S* coupling [23] are applicable. For triplet series we compare in Table 3 theoretical results obtained according to formulas for triplet case [25] with the sum over three experimental components, which were estimated from Fig. 2 of Ref. [1].

The satellite intensities relative to Ar $1s^{-1}$ line were calculated using formulas (1) and (3). The results corrected for the photoionization cross-section according to formula (6) are also presented there. The intensity of the satellite, corresponding to excitation 3p→4p is in good agreement with experiment and theoretical data in sudden approximation [12]. It is seen from Table 3, that the satellite intensities for the excitation 2p→4p are also in good agreement with experiment [1]. The corrections by photoionization cross-section are not significant in these two cases. The Coulomb matrix elements for the excitations 3p→4p and 2p→4p differ insignificantly and the ratio of intensities is almost exactly equal to the squared ratio of the excitation energies.

The suggestion in [1] that the state $1s^{-1}2p^{-1}4s$ in the singlet channel is created as a shake-up excitation 1s→4s, accompanying the 2*p*- ionization, results in a very small value of intensity (see Table 3). There are several reasons for this. Squared energy denominator in (4) results in a factor $10^{-2}$ relative to the $1s^{-1}2p^{-1}4p$ satellite state. Direct and exchange Coulomb integrals have different signs in this case. Note that the signs of weight factors before Coulomb and exchange integrals are the same in a singlet case and are different in a triplet case. In the case of triplet 4*s* satellite negative signs of exchange integral and exchange contribution to the weight factor result in the same signs for direct and exchange terms. That is why the theoretical intensity of triplet 4*s* satellite is two orders of magnitude larger then the intensity of singlet 4*s* satellite. Even after correction of Ar 2*p* photoionization cross-section for the energy of satellite excitation 1s→4s (see Table 1) the theoretical intensity of singlet 4*s* state is two orders of magnitude less than the experimental value (see Table 3).

Table 3. Theoretical and experimental shake-up intensities in $10^{-4}$ of Ar 1*s* photoelectron line

| Final state | Formula (3) | Formula (6) | Exp. [1, 10] |
|---|---|---|---|
| $1s^{-1}3p^{-1}4p(^2S)$ | 680[a] | 680 | ≈600 |
| $1s^{-1}2p^{-1}(^1P_1)4p(^2S)$ | 2.6 | 3.0 | ≈2.75 |
| $5p(^2S)$ | 1.0 | 1.18 | ≈0.75 |
| $6p(^2S)$ | 0.48 | 0.57 | ≈0.37 |
| $1s^{-1}2p^{-1}(^3P_{0,1,2})4p(^2S)$ | 4.5 | 5.3 | ≈2.08 |
| $5p(^2S)$ | 1.7 | 2.0 | ≈0.79 |
| $1s^{-1}2p^{-1}(^1P_1)4s(^2P)$ | $0.30 \cdot 10^{-5}$ [b] | $0.60 \cdot 10^{-3}$ [c] | ≈0.33 |
| $1s^{-1}2p^{-1}(^3P_{0,1,2})4s(^2P)$ | $0.96 \cdot 10^{-3}$ [b] | 0.19 [c] | ≈0.23 |

a) The sum of $1s^{-1}3p^{-1}(^1P)4p$ and $1s^{-1}3p^{-1}(^3P)4p$ states, whose theoretical intensities equal to 0.025 and 0.043 of Ar 1s line intensity, respectively
b) Spectroscopic factors were multiplied by the ratio of photoionization cross-sections $\sigma_{2p}/\sigma_{1s} = 0.0383$ (photon energy 3900 eV for both shells).
c) Spectroscopic factors were multiplied by the ratio of photoionization cross-sections $\sigma_{2p}/\sigma_{1s} = 7.63$ (photon energy 700 eV for 2p–shell and photon energy 3900 eV for 1s-shell).

The states corresponding to monopole shake-up excitations $1s \rightarrow 4s$ accompanying $2p$ ionization may be also obtained by dipole direct knock-up processes $2p \rightarrow 4s$ accompanying $1s$ photoionization. Dipole interaction of outgoing electron with ionic core may also result in excitation $2p \rightarrow 3d$. Theoretical energies of $4s$ and $3d$ states in DCH potential and intensities of direct knock-up excitations are presented in Table 4.

Total non-relativistic energies were calculated self-consistently for all terms. Since we compare relative energies for singlet and triplet states separately, non-relativistic approach is sufficient in this case. The intensities of direct knock-up processes were calculated using the formula (5) for actual coupling scheme. The calculation demonstrated that the energy differences between $4s$ state and some terms of $3d$ states are less then the linewidth of the satellites 0.6 eV [1] (see Table 4). The largest direct knock-up intensity of the state $1s^{-1}2p^1(^1P)3d(^2P)$ in a singlet channel is 0.412 and the energy shift from the state $1s^{-1}2p^1(^1P)4s(^2P)$ is 0.45 eV. Hence it follows that the experimental intensity of the state labeled $4s$ in a singlet channel [1] is due to direct knock-up to the $3d$ state.

On the other hand, in a triplet channel the energy splitting between $4s$ state and the only state $3d(^2P)$ of significant intensity is 1.09 eV, i.e. larger than the experimental line width. Hence, it follows that the intensity of $4s$ satellite in triplet channel cannot be explained by direct knock-up process. At the photon energy 3900 eV the photoionization cross section of $2p$ shell is 26 times less than that of $1s$ shell and theoretical intensity of $4s$ satellite in triplet channel without correction for the energy transfer is too small (see Table 3). The $2p$ photoionization cross-section corrected for the energy transfer to the satellite excitation becomes 7.6 times larger than that of $1s$ shell (see Table 1) and $4s$ triplet satellite intensity is in agreement with the experiment (see Table 3). Thus, in order to explain the intensities of some satellite states in DCH field corrected photoionization cross-section should be used and direct knock-up process should be taken into account.

Table 4. Intensities of direct knock-up satellites of Ar 1s photoionization (in units of $10^{-4}$ of Ar1s line) and binding energies relative to the state $1s^{-1}2p^{-1}(^1P)4s(^2P)$

| Final state | $\Delta E_{b(eV)}$ | Intensity | Final state | $\Delta E_{b(eV)}$ | Intensity |
|---|---|---|---|---|---|
| $1s^{-1}2p^{-1}(^3P)4s(^2P)$ | 0.00 | 0.00056 | $1s^{-1}2p^{-1}(^1P)4s(^2P)$ | 0.0 | 0.0034 |
| $1s^{-1}2p^{-1}(^3P)3d(^2P)$ | -1.09 | 0.579 | $1s^{-1}2p^{-1}(^1P)3d(^2P)$ | -0.45 | 0.412 |
| $1s^{-1}2p^{-1}(^3P)3d(^2D)$ | -0.87 | 0.0060 | $1s^{-1}2p^{-1}(^1P)3d(^2D)$ | -0.87 | 0.0020 |
| $1s^{-1}2p^{-1}(^3P)3d(^2F)$ | -0.37 | 0.0016 | $1s^{-1}2p^{-1}(^1P)3d(^2F)$ | -0.34 | 0.113 |
| Total, theory | | 0.587 | Total, theory | | 0.530 |
| Experiment [1] | | $\approx 0.23$ | Experiment [1] | | $\approx 0.33$ |

## IV. CONCLUSIONS

High resolution technique in HARPES region [1] permits to resolve the fine structure of previously known [2] shake-up excitations $2p \to 4p$ of Ar 1s- ionization and to identify new and very small ($10^{-4}$ of the main line) excitations, whose symmetry corresponds to $1s \to 4s$ satellites of Ar 2p-ionization. The latter ones present a new type of electron correlation satellites in photoelectron spectroscopy with extremely large (about 3000 eV) energy parameter. The results of our calculation using standard approaches of many-body-perturbation theory [25] are in quantitative agreement with the experiment [1] and confirm the shake-up mechanism of the excitations from 2p-shell, accompanying the Ar 1s ionization.

The smallest lines, corresponding to $1s \to 4s$ excitations and identified for the first time in [1] are the subject of reconsideration of the many-body shake-up theory. It was shown in this paper that in order to explain intensities of the satellite excitations $1s \to 4s$ in triplet channel, the photoionization cross-sections with account for the energy transfer to satellite has to be used [25]. In the case under consideration it corresponds to the factor up to about 200. It was also obtained that the energy positions of 3d and 4s states in DCH potential are quite close to each other and that the intensities of some direct knock-up processes $2p \to 3d$ are of the same order as experimental satellite intensities in 4s channel. Thus, the origin of the smallest lines in DCH spectra may most probable due to direct knock-up process. It means that the measurements performed in [1] permit to obtain and identify not only the strong lines but gain new important information from the low intensity features of the measured spectrum.

1. R. Püttner, G. Goldsztejn, D. Céolin, J.-P. Rueff, T. Moreno, R. K. Kushawaha, T. Marchenko, R. Guillemin, L. Journel, D. W. Lindle, M. N. Piancastelli, and M. Simon. Phys. Rev. Lett. **114**, 093001 (2015).
2. U. Kuetgens and J. Hormes Phys. Rev. A **44**, 264 (1991)
3. C. S. Fadley, *Hard X-ray Photoemission: An Overview and Future Perspective*. In *Hard X-ray Photoelectron Spectroscopy (HAXPES)* Springer Series in Surface Sciences, **59,** 1-34 (2016).
4. M. Simon, M. N. Piancastelli and D. W. Lindle, *Hard-X-ray Photoelectron Spectroscopy of Atoms and Molecules*. In *Hard X-ray Photoelectron Spectroscopy (HAXPES)* Springer Series in Surface Sciences, **59**, 65-110 (2016).
5. V. I. Nefedov, V. G. Yarzhemsky, R. Hesse, P. Streubel, and R. Szargan, J. Electron Spectrosc. Relat. Phenom., **125**, 153 (2002).
6. M. Ya. Amusia, L. V. Chernysheva, and V. G. Yarzhemsky, JETP Letters, **97**, 704 (2013).
7. S. Svensson, B .Eriksson, N. Martensson, G. Wendin, and U. Gelius, J. Electron Spectrosc. Relat. Phenom., **47**, 327 (1988).
8. M. Y. Adam, F. Wuilleumier, S. Krummacher, V. Schmidt, and W. Mehlhorn, J. Phys. B: Atom. Mol. Phys., **11**, L413 (1978).
9. M. O. Krause, S. B. Whitfield, C. D. Caldwell, J.-Z. Wu, P. van der Meulen, C. A. de Lange, and R. W. C. Hansen, J. Electron Spectrosc. Relat. Phenom., **58**, 79 (1992).
10. M. Pahler, C. D. Caldwell, S. J. Schaphorst, and M. O. Krause, J. Phys. B: At. Mol. Opt. Phys. **26**, 1617 (1993).
11. A. Kikas, S. J. Osborne, A. Ausmees, S. Svensson, O.-P. Sairanen, and S. Aksela, J. Electron Spectrosc. Relat. Phenom., **77**, 241 (1996).
12. S.H. Southworth, T. LeBrun, Y. Azuma, and K. G. Dyal, J. Electron Spectrosc. Relat. Phenom., 94, 33 (1998).
13. V. G. Yarzhemsky, M. Ya. Amusia, P.Bolognesi, L. Avaldi, J. Phys.B:Atomic Molec Opt. Phys., **43**, 185204 (2010).
14. T. Aberg, Phys. Rev. **156**, 35 (1967).
15. R. L. Martin and D. A. Shirley, Phys. Rev. A 13, 1475 (1976).
16. K. G. Dyall and F. P. Larkins, J. Phys. B: At. Mol. Phys., 15, 1021 (1982).
17. K. G. Dyall, J. Phys. B: At. Mol. Phys., 16, 3137 (1983).
18. B. M. Lagutin, I. D. Petrov, V. L. Sukhorukov, S. B. Whitfield, B. Langer, J. Viefhaus, R. Wehlitz, N. Berrah, W. Mahler, and U. Becker, J. Phys. B: At. Mol. Opt. Phys., 29, 937 (1996).
19. V. G. Yarzhemsky, G. B. Armen, and F. B. Larkins, J. Phys. B: At. Mol. Opt. Phys., 26, 2785 (1993).
20. A. S.Kheifets, J. Phys. B: At. Mol. Opt. Phys., 29, 3791 (1995).
21. T. Kaneyasu, Y. Hikosaka, E. Shigemasa, F. Penent, P. Lablanquie, T. Aoto, and K. Ito, J. Phys. B: At. Mol. Opt. Phys., **40**, 4047 (2007).
22. V. G. Yarzhemsky, A. S. Kheifets, G. B. Armen, F. P. Larkins, J. Phys. B: At. Mol. Opt. Phys. **28**, 2105 (1995).
23. V. G .Yarzhemsky, F. P. Larkins, The European Physical Journal D - Atomic, Molecular, Optical and Plasma Physics, **5**, 2, 179-184 (1999).
24. V. G. Yarzhemsky, M. Ya. Amusia, and L. V. Chernysheva, J. Electron Spectrosc. Related Phenom., **127**, 153 (2002).

25. M. Ya. Amusia, L. V. Chernysheva and V. G. Yarzhemsky, *Handbook of Theoretical Atomic Physics*, Data for Photon Absorption, Electron Scattering, and Vacancies Decay, Springer-Verlag, Berlin, Heidelberg, 2012.
26. M. Ya. Amusia and L. V. Chernysheva, *Computation of atomic processes*, IOP Publishing Ltd: Bristol and Philadelphia, 1997.
27. I. Lindgren and J. Morrison, *Atomic many-body theory*, Springer, Berlin, 1982
28. B. R. Judd, *Operator Techniques in Atomic spectroscopy*, McGraw-Hill, 1963.
29. M. B. Trzhaskovskaya, V. I. Nefedov, V. G. Yarzhemsky, At. Data Nucl. Data Tables, **82**, 257 (2002).
30. L. V. Chernysheva and V. L. Yakhontov, Computer Phys. Communications, **119**, 232 (1999)